# An Objectives-Driven Process for Selecting Methods to Support Requirements Engineering Activities

Lester O. Lobo and James D. Arthur

Department of Computer Science, Virginia Tech

Blacksburg, Virginia  24060, USA

{lester, arthur}@vt.edu

**ABSTRACT.**  This paper presents a framework that guides the requirements engineer in the implementation and execution of an effective requirements generation process. We achieve this goal by providing a well-defined requirements engineering model and a criteria based process for optimizing method selection for attendant activities. Current models address only portions of the requirements process or include activities defined at higher levels of abstraction; this often obscures the implementation aspects of the requirements process. Our model, unlike other models, addresses the complete requirements generation process and consists of activities defined at more adequate levels of abstraction. Additionally, activity objectives are identified and explicitly stated - not implied as in the current models. Activity objectives are crucial as they drive the selection of methods for each activity. Thus, our model guides the requirements engineer through the requirements generation process by providing a clear transition path for requirements through adequately decomposed, well-defined activities. Our model also incorporates a unique approach to verification and validation that enhances quality and reduces the cost of generating requirements. To assist in the selection of methods, we have mapped commonly used methods to activities based on their objectives. In addition, we have identified method selection criteria and prescribed a reduced set of methods that optimize these criteria for each activity defined by our requirements generation process. Thus, the defined approach assists in the task of selecting methods by using selection criteria to reduce a large collection of potential methods to a smaller, manageable set. The model and the set of methods, taken together, provide the much needed guidance for the effective implementation and execution of the requirements generation process.

## 1. INTRODUCTION

In the past two decades much has been done towards improving the software development process, with the goal of improving the project success rate. However, according to the Standish report, only 28% of the real world projects are successful [1].

Williams attributes this low rate of success primarily to the lack of clear and precise requirements [2]; the reason being that a system is only as good as the requirements from which it is developed. The statistics reported by the Standish group indicate that the software industry still lacks an effective definition of the requirements generation process. This is because the requirements engineers lack guidance in two different areas – implementing the requirements engineering model, and selecting methods for activities within the model.

The first problem is deried from the observation that many models do not adequately address the requirements generation process. Current requirements engineering models such as the Win-Win model [3] and Requirements Triage [4] either have a narrow focus on only portions of the requirements generation process, or provide a broader perspective defined by abstract, high-level activities. Additionally, the models often include implicit activity objectives. That is, they lack a clear mapping of objectives to activities. Consequently, the requirements engineers may ignore the implied objectives which can adversely impact the project's success. Thus, implementing these models is difficult because they provide an obscured/narrow view of a complete and sufficiently decomposed requirements engineering process.

The next hurdle that a requirements engineer faces is selecting methods for the activities in the requirements engineering model. It is widely acknowledged that methods have significant impact on the quality of the final product [5]. Hence, a substantial amount of research has been conducted in identifying methods for the entire software development life cycle. As a result, there are a large number of methods for the requirements engineering process. To date, however, these methods are mapped to the high-level, abstract activities (e.g. elicitation, analysis, specification); there is a noticeable lack of the coordination of methods with lower level activities. [6]. Given this scenario, the requirements engineer often selects methods in an ad-hoc fashion, resulting in an output which insufficiently addresses the objectives of activities in the requirements model.

This paper describes a framework that guides the requirements engineer in two critical tasks: (1) implementing an effective requirements generation model, and (2) selecting methods for the various requirement engineering activities. To provide guidance in the implementation of the requirements process, we propose a two-phase model that is well-defined and which addresses the entire requirements generation process. In addition, the model consists of activities decomposed at an adequate level of granularity to facilitate the selection of methods. This decomposition also provides a seamless evolutionary path for requirements as they move through successive activities. Because activity objectives drive the selection of methods, the objectives of each activity in the model are identified and *explicitly* stated, that is, there are no implied objectives. An added advantage of the model is that it includes a unique approach to verification and validation (V&V) that enhances the quality of the requirements generated, and reduces the time and effort associated with the overall V&V activities. To assist the requirements engineer in the second critical task of choosing the appropriate methods for a particular activity, we identify methods commonly used in the industry and map them to the decomposed activities, based on their stated objectives. We provide additional guidance in the selection of methods by grouping them according to criteria they purport to optimize, e.g., cost, personnel, time, or completeness. Thus, given a selection criteria, the requirements engineer can easily trace a path of methods (within the reduced set) that optimize the chosen criteria for the entire requirements generation process.

The subsequent sections in this paper further describe our approach to providing guidance to the requirements engineer. In Section 2, we describe the decomposition of the two-phase model into its constituent activities and discuss them in detail. Section 3 identifies the method selection criteria and discusses the process of choosing the sequence of methods to optimize the chosen selection criteria. Finally, Section 4 presents the summary and possible future work.

## 2. DECOMPOSITION OF THE REQUIREMENTS GENERATION MODEL

Our model extends the Requirements Generation Model [7] by decomposing its relatively high-level set of activities into a more definitive set with explicitly identified objectives.

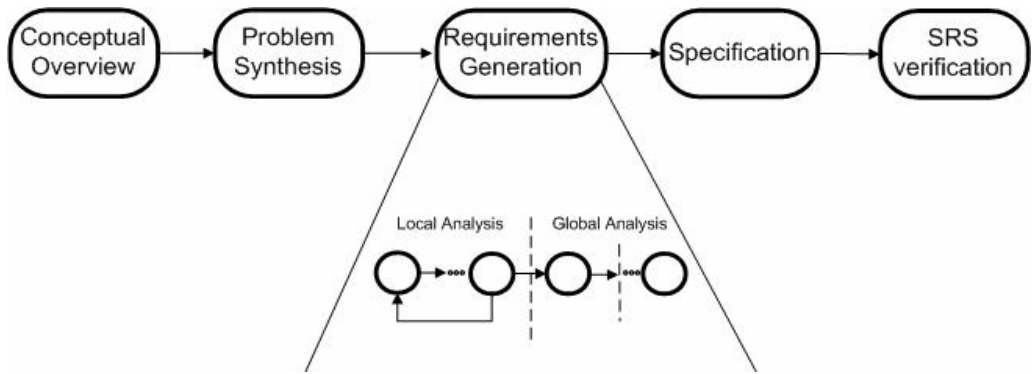

**Figure 1**: Requirements Generation Model

The five phases in the original RGM are:

- *Conceptual overview*: helps recognize the need for a new system from the business and operational perspectives.

- *Problem Synthesis*: assists identifying customer problems and needs.

- *Requirements generation*: involves elicitation, analysis, and production of quality adherent requirements.

- *Specification*: helps in the documentation of requirements into the formal SRS.

- *SRS verification*: Final verification of the SRS before submitting for customer's approval.

Our focus is on the critical Requirements Generation phase because this phase transforms the customer needs into concrete software requirements. We have decomposed the Requirements Generation phase into activities based on the concept of "Separation of Operations" which focuses on identifying and satisfying a small set of concerns for organizing and decomposing complex process [8]. Our primary concerns during the decomposition of the model were: (1) the activities should have a single, focused objective and (2) the decomposition should facilitate the selection of methods for activities based on their objectives.

Our model integrates seamlessly into the requirements engineering life cycle because it uses the RGM as an extendible basis. We have decomposed the Requirements Generation phase into two sub-phases:

- *Local Analysis*: an iterative phase focusing on eliciting, analyzing, documenting and evaluating incremental sets of requirements, and

- *Global Analysis*: which complements the Local analysis phase, and concentrates on selected verification and related business concerns of the more comprehensive set of requirements.

Our proposed model emphasizes early verification and validation of requirements to overcome the drawbacks of current models that accentuate the V&V activities late in the requirements life cycle – immediately prior to the generation of the formal requirements document (SRS). As a result, the V&V activities in these models are burdened with the evaluation of one whole set of requirements at one time which makes it difficult, if not impossible, to focus on the quality of individual requirements. In addition, because the V&V activities are conducted towards the end of the requirements phase, and far removed from the elicitation of requirements, an additional amount of effort must be expended by the stakeholders to revisit the requirements, their context and rationale. Often in this process, there is a loss of information because the requirements become obscured in the minds of the stakeholders because of the large time delay between the requirements elicitation and V&V activities. Our model, consisting of Local and Global Analysis phases, focuses on early V&V to alleviate these problems associated with software development. The Local Analysis phase brings V&V closer to the Elicitation activity by performing V&V on smaller, incremental sets of requirements – not the complete set. On the other hand, the Global Analysis phase focuses on verifying the linkages between sets of requirements rather than within the sets. The effect of such an approach is that it results in early error detection and correction of requirements. Consequently, the cost incurred during project development is minimized and the quality of the requirements is enhanced.

The detailed explanation of these phases is provided in the subsequent sections.

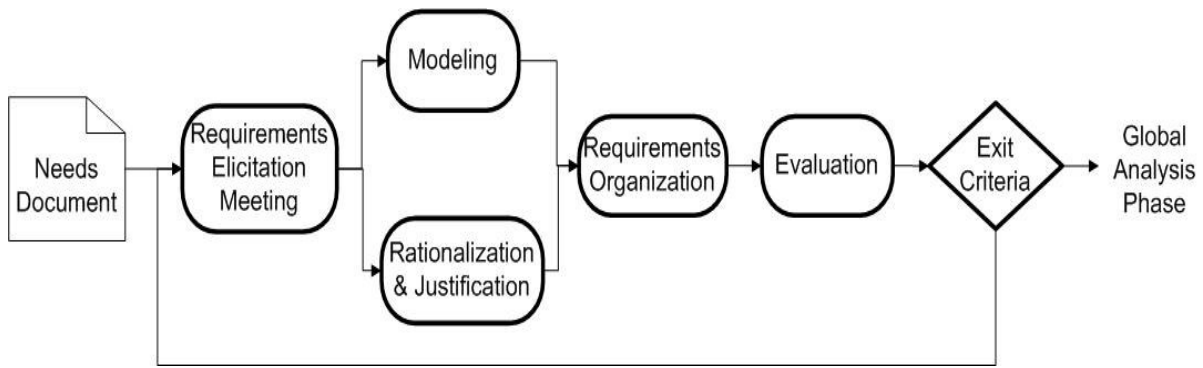

**Figure 2**: Decomposition of the Local Analysis phase

## 2.1 Local Analysis Phase

Local Analysis is an iterative phase through which the customer and requirements engineer discover, review, articulate and evaluate *incremental* sets of requirements corresponding to *functional partitions* of the system. Figure 2 shows the decomposition of the Local Analysis phase into its constituent activities.

The Needs Document, generated by the Problem Synthesis phase, is the input to the first activity – Requirements Elicitation Meeting. The objective of this activity is to correctly identify and capture requirements of the stakeholders. During this activity, the responsibilities of the stakeholder and the requirements engineer are complementary – one provides information, the other captures information.

The next two activities help in understanding and analyzing the elicited requirements. Modeling provides a clear representation of the requirements and helps in better evaluating the requirements [9]. Additionally, the generated models assist in easier validation of the requirements. They are also useful during the Global Analysis phase when assessing the impact of business/organizational concerns. Furthermore, the models are also available for the design phase.

We have included a Rationalization and Justification activity for identifying the rationale behind the elicited requirements. This is helpful because stakeholders are often vague in the description of their requirements. An analysis of the rationale helps justify the existence of the requirements in the specification. The Rationalization and Justification activity also enables tracing the requirements back to the needs through an examination of the requirements rationale/reasoning.

As a product of Modeling, and Rationalization and Justification activities, the requirements are represented as unordered lists. Hence, to provide structure to the requirements we included the Requirements Organization activity. This activity also helps identify the important requirement attributes such as risk factors, customer importance, effort required, and so forth. The organization of requirements involves hierarchically classifying the requirements on a functional and non-functional basis.

The Evaluation activity includes requirements verification, validation, and conflict resolution as related to the functional partitions of the system. The objective of verification is to determine whether the requirement adhere to quality characteristics such as non-ambiguity, correctness, verifiability, and the like. Some of the quality attributes such as completeness, traceability, and consistency are only partially evaluated because total evaluation requires the availability of the complete set of requirements. Validation assists in determining whether the requirements satisfy the customer intent. Like verification, validation is also performed on incremental sets of requirements. Inconsistencies identified during the V&V are then resolved through the effective interest based bargaining approach [10].

To determine when to exit the iterative Local Analysis phase, we examine whether or not the requirements satisfy a checklist of exit criteria. The necessary items in this list pertain to: (a) inspecting requirements quality attributes, (b) ensuring the requirements necessarily and sufficiently trace back to the needs and (c) finding agreement among stakeholders that all requirements have been elicited.

## 2.2 Global Analysis Phase

The Global Analysis Phase focuses on the complete set of requirements that is generated by the preceding Local Analysis phase. This phase includes two sub-components (illustrated in Figure 3):

- *Global Evaluation*: completes the verification process and resolves conflicts, and
- *Business Concerns*: assists in the evaluation of requirements from the business perspective.

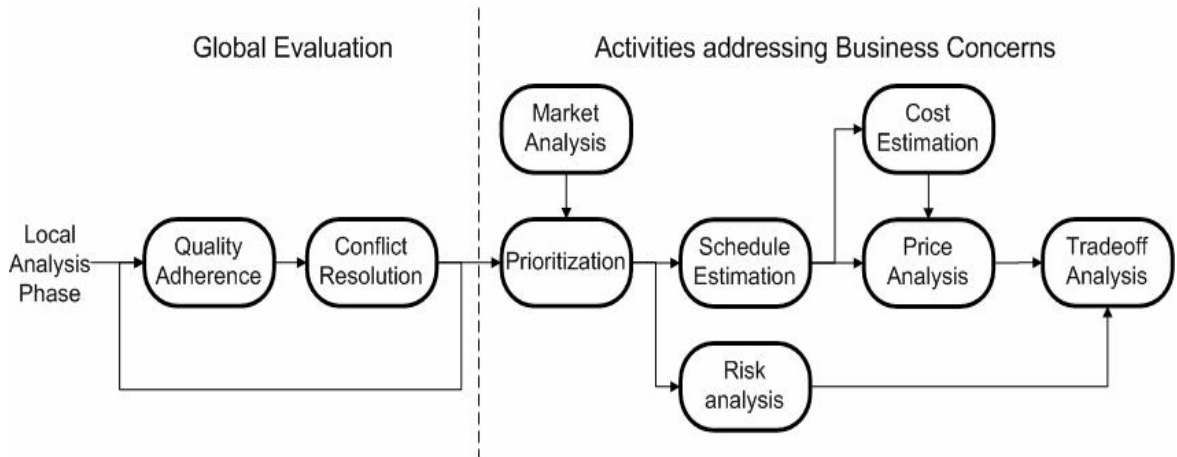

**Figure 3**: Decomposition of the Global Analysis Phase

The Global Evaluation component consists of Quality Adherence and Conflict Resolution activities. The Quality Adherence activity complements the verification process in the preceding Local Analysis phase. Its focus is on the quality attributes – completeness, traceability and consistency. During the Quality Adherence activity, effort is expended on analyzing the quality of linkages between the requirements sets – not within the sets. Validation is optional in the Global Evaluation component because it has been previously performed in the Local Analysis phase. However, if the customer desires, validation can be performed as a proof of concept. Inconsistencies identified during verification are addressed in the Conflict Resolution activity in a non-confrontationist atmosphere.

The Business concerns component helps determine project feasibility and scope, and helps address organizational issues and constraints. The activities in this component are an extension to Requirements Triage model proposed by Alan Davis [11]. However, unlike the Triage model which has a narrow focus, our approach covers the entire requirements process and provides for a seamless integration of the Business Concerns component with the rest of the model.

A brief description of the activities in this component is provided below:
- *Market Analysis*: helps collect market information such as user expectation, market trends, competitor's product features, and so forth.

- *Prioritization*: assists in ranking requirements based on importance to the user and the value added to the product.
- *Schedule Estimation*: useful in determining the development time and identifying critical software components.
- *Risk Analysis*: focuses on examining the requirements for risk factors pertaining to product engineering, development environment and program constraints [12].
- *Cost Estimation*: helps in determining the development effort and hence, cost required in building the system.
- *Price Analysis*: assists in reaching a fair and reasonable price for the product independent of the cost of individual components and proposed profit.
- *Tradeoff Analysis*: helps evaluate the pros and cons of the system in Operational, Technical, Schedule, Economic, and Legal terms. It also focuses on resolving conflicts and determining the scope of the system.

Thus, on completion of the Global Analysis phase, we obtain complete set of requirements that satisfy the customer needs and which have a well-defined scope. A more detailed discussion of the model can be found in [13]

## 2.3   Identifying Methods for Activities

In order to map methods/techniques to requirements engineering activities, it is crucial to determine the objectives of each activity; this is because activity objectives drive the selection of methods. Hence, one of our primary concerns during the decomposition process was to identify activities which have focused and explicit objectives. Most of the current models such as RE Process Framework [14], and Win-Win model [15] provide a high-level perspective of the requirements generation process. Hence, while higher level objectives are explicitly stated lower level ones are often implied or ignored. For example, modeling and identifying the rationale are often implied objectives of requirements analysis. Thus, within the expanded RGM model we have attempted to define activities at an adequate level of abstraction supporting the enunciation of clear, explicit objectives.

Once the activity objectives are defined, the next step is to identify methods that closely satisfy the activity objectives. Our initial goal was to identify all possible techniques for the various activities in the model. We decided, however, to focus on methods commonly used in the industry. This was because the literature included a large number of methods and only a fraction of those were actually being employed in the industry. Additionally, the goal of our research was to provide guidance to the requirements engineer in the "real world". The mapping of methods to activities is illustrated for two activities in Table 1.

| Activity Name | Activity Objective | Applicable Methods |
|---|---|---|
| Requirements Elicitation Meeting | Correctly identify and capture requirements from the stakeholders | Interviews, Observation, Task Demonstration, Document Studies, Questionnaires, Brainstorming, Focus Groups, Requirements Workshops, Prototyping |
| Risk Analysis | Estimate risk in the development of system components | Criticality Analysis, Fault Tree Analysis, Risk Reduction Leverage, Event Tree Analysis, Monte Carlo Simulation, FMECA (Failure mode, effects, and criticality analysis) |

**Table 1**: Mapping of methods to activities based on their objectives

The complete mapping of methods to all activities in our proposed model is provided in [13]. A total of 77 methods have been identified for the 14 activities in the expanded RGM. As a result of this mapping we achieve two goals. First, we have synchronized effective methods with well-defined and appropriately decomposed activities. In comparison, the requirements engineering literature identifies a large number of methods for high-level activities with no clear mapping between methods and objectives. Secondly, we provide a reduced set of methods for the activities and thus, make the task of selecting appropriate methods for an activity easier.

### 2.4 Advantages of the Proposed Model

Several benefits are apparent in our proposed approach; they are outlined below:

1) ***Well-defined model***: Our model addresses the entire requirements generation process and includes activities at appropriately decomposed levels of granularity. This is in contrast with other current models which either have a narrow focus or provide only high level perspectives of the requirements engineering process.

2) ***Explicit objectives***: The selection of methods for an activity is driven by its objective(s). Hence, our decomposition process identifies activities that have clear, focused objectives. Every activity objective is stated explicitly - not implicitly, which is prevalent in many of the current models because of their high level focus on the requirements generation process.

3) ***Synchronization of methods with activities***: We have mapped the commonly used methods in the industry to the activities based on their objectives. Additionally, unlike previous method mapping research, our work is based on an appropriate level of activity decomposition illustrating a clear synchronization between methods and activity objectives.

4) ***Easier method selection***: Because we provide a smaller set of methods for each activity, it is easier for the requirements engineer to select an appropriate set of overall methods.

Even though we have identified a reduced set of methods for the requirements generation process, the number of methods for each activity is fairly large. From Table 1, we can see that there are nine methods to choose from for the Elicitation activity and six for the Risk Analysis activity. In a real world scenario, the requirements engineer makes his/her choice based on certain method selection criteria and by weighing the strengths and weakness of each method against that criterion. Our goal, as detailed in the next section, is to simplify this task.

## 3. CRITERIA BASED METHOD SELECTION

A substantial part of our research has been to identify an appropriate set of methods that support individual requirements engineering activities and the achievement of their stated objectives. Selecting the *one* method within that set, however, is often based on operational or organizational criteria that have little bearing on the activity objectives. For example, we might desire to select the method that has the least cost to implement. To provide guidance in the selection process, we have chosen four criteria commonly used in the industry, and have analyzed each method relative to its ability to support the achievement of that criteria. The four criteria are introduced next, followed by method analyses and the selection process.

- *Personnel*: Selection based on the number of people involved and their expertise. This criterion is the most widely used criterion in the selection of methods because a project usually has limited work staff.

- *Time*: Selection based on the time needed to perform a technique. Often the requirements engineer has very limited time to perform a particular activity. In such a situation, it is necessary to employ the technique which achieves the activity objective in the least amount of time.

- *Cost*: Selection based on expenses involved in conducting the method. In situations where the project is under budgeted, it is imperative to select techniques that minimize the cost incurred.

- *Completeness*: Selection based on the coverage of activity objectives. This criterion is used when it is necessary to completely achieve the objective of one activity before proceeding to the next activity, e.g. life critical systems.

For each of the above criteria, we identified the methods for each activity in the expanded RGM that best optimize each selection criteria. This results in a smaller set of methods for each activity, which in turn simplifies the selection task of the requirements engineer. The concept of using the selection criteria in choosing methods for an activity is illustrated in Figure 4. The bubbles $C_1, C_2, C_3$ and $C_4$ represent the set of methods which satisfy each of the four selection criteria. For example, $C_1$ may represent methods optimizing cost criteria, $C_2$ – methods satisfying completeness criterion, and so forth. The three activities depict three different scenarios in which the set of methods accommodates the four selection criteria:

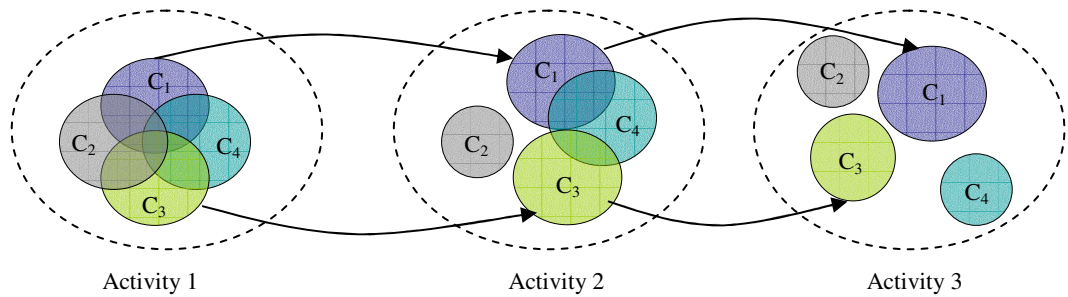

**Figure 4**: Use of selection criteria in choosing methods

- *Ideal scenario*: This occurs when there are a set of methods which satisfy all the four criteria as shown by the intersection of the four bubbles in Activity 1. In this situation, the requirements engineer's task of method selection is the easiest with the selected method(s) satisfying *all criteria*.

- *Normal scenario*: Activity 2 illustrates the normal scenario where there exist a set of methods which satisfy some criteria but not all. As seen from Activity 2, C4 intersects with $C_1$ and $C_3$, but not with $C_2$. Also, $C_2$ does not intersect with any of the other bubbles. Hence, in such situations, the choices of methods often satisfy *some criteria* but compromise on the others.

- *Worst case scenario*: This occurs when the methods satisfy different criteria as shown in Activity 3. Consequently, the requirements engineer must select methods that satisfy *one criterion* while compromising the rest.

Thus, given the framework where the methods supporting a single activity are grouped based on selection criteria, the task of the requirements engineer is simplified because s/he can select from a reduced set of methods that optimize the selection criteria. Suppose the requirements engineer needs to identify methods that optimize the cost criterion. If in Figure 4 the bubble $C_1$ represents methods satisfying cost criterion, then the requirements engineer has to only examine those methods in bubble $C_1$ for each of the three activities in order to select the appropriate methods. The arrows in the figure illustrate this selection process. Additionally, within the set of methods satisfying the chosen criterion, the methods can be selected by comparing them based on their documented strengths and weaknesses. Furthermore, this framework and approach enables the requirements engineer to achieve better results by facilitating the selection of methods that satisfy multiple criteria. Another advantage is that the number of methods employed can be minimized by selecting a minimal set of methods that meet the objectives of several activities. Thus, our framework simplifies the task of selecting a path of methods for the entire requirements generation process.

## 3.1 A Practical Example

In this section we illustrate the selection process and the usefulness of the method selection criteria. Here we consider the Risk Analysis activity which assists in determining the risk involved in the development of the software components. Based on the literature and industry practice, we identified six risk analysis methods that are widely used in software project development. These methods were then analyzed to determine those which best optimize cost, time, personnel, and completeness. The mapping of methods to the Risk Analysis activity based on the four selection criteria is depicted below.

| Risk Analysis Activity | |
|---|---|
| **Selection Criteria** | **Methods for the Risk Analysis activity** |
| Personnel | FMECA (Failure mode, effects, and criticality analysis) [16], Monte Carlo Simulation |
| Time | Criticality Analysis [MILL 77], Monte Carlo Simulation |
| Cost | FMECA (Failure mode, effects, and criticality analysis), Criticality Analysis |
| Completeness | Monte Carlo Simulation, Fault Tree Analysis and Event Tree Analysis [17] |

**Table 2**: Methods optimizing selection criteria for the Risk Analysis activity

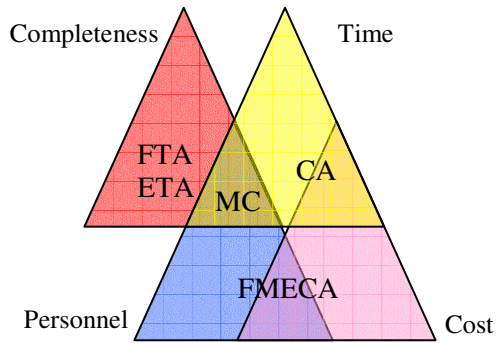

| MC | Monte Carlo Simulation |
|---|---|
| CA | Criticality analysis |
| FTA/ETA | Fault Tree Analysis and Event Tree Analysis |
| FMECA | Failure mode, effects, and criticality analysis |

| Criteria | Methods for Risk Analysis Activity | | | |
|---|---|---|---|---|
| | FMECA | MC | CA | FTA & ETA |
| Personnel | X | X | | |
| Time | | X | X | |
| Cost | X | | X | |
| Completeness | | X | | X |

**Figure 5:** Methods optimizing multiple criteria for the Risk Analysis activity

As seen from Figure 5, Monte Carlo Simulation technique optimizes three selection criteria – time, personnel and completeness. Hence, this technique is the preferred method because it optimizes three selection criteria. However, this does not imply that Monte Carlo Simulation is the best for all the three criteria taken separately. In fact, if time is the deciding factor and other criteria are marginalized, Criticality Analysis is a better choice than Monte Carlo Simulation. This can be judged by analyzing the pros and cons of both these methods. The method, Criticality Analysis, optimizes both the time and cost criteria. In situations where there is a primary and secondary criterion, such as time and cost, Criticality Analysis is the most suitable method.

Another consideration in the selection of methods is to employ a minimal set of methods for the requirements generation process. Our framework facilitates this goal by identifying methods that optimize the selection criteria for all activities in the requirements process and thus, enabling the requirements engineer to take informed decisions. For example, if interviews is the preferred method for the Elicitation activity and is the second preference for the Rationalization and Justification activity, the requirements engineer may select interviews for the latter activity to reduce the overhead of performing two different methods. Thus, from the discussed example, we see that the method selection task of the requirements engineer is simplified through two features – (1) documentation of the pros and cons of the methods (2) grouping of methods based on selection criteria.

Our framework facilitates the selection of a path of methods that optimize the chosen selection criteria for the entire requirements generation process. This is illustrated in Table 3 which maps methods to activities in the Business Concerns component, starting with the Risk Analysis activity, based on the completeness criterion.

| Completeness Criteria Applied to Activities of the Business Concerns Component | | | | |
|---|---|---|---|---|
| **Risk Analysis** | **Cost Estimation** | **Schedule Estimation** | **Price Analysis** | **Tradeoff Analysis** |
| • Monte Carlo Simulation<br>• Criticality Analysis | • COCOMO II<br>• Function Point Approach | • PERT (Program Evaluation and Review Technique)<br>• CPM (Critical Path Method) | • Comparative Price Analysis<br>• Value Analysis | • PMI (Plus, Minus, and Implications)<br>• Decision Analysis<br>• Internal Rate of Return<br>• Net Present Value |

**Table 3**: Mapping of methods based on completeness criterion

As depicted in the above table, there are a small set of methods for each activity. The choice among these methods can be made by studying the documented strengths and weaknesses of the methods. This approach certainly has an advantage over the current situation where the requirements engineer is provided with a large collection of methods mapped to high level activities having implicit objectives. A sample path of methods for the activities in Table 3 based on completeness criterion could be:

> **Monte Carlo Simulation**
> →  COCOMO II
>    →  PERT
>       →  Comparative Price Analysis
>          →  Net Present Value

Such a table is provided for each of the four criteria which facilitate the making of informed decisions on method selection for a particular activity. Based on our research we can, on the average, filter out nearly one-half of 77 methods for each selection criteria in the requirements generation process. Furthermore, for each activity the choice is then typically reduced to 2-3 methods. This enables the requirements engineer to focus on selecting the most appropriate method that not only satisfies the project constraints but also the activity objectives.

Thus, our proposed framework provides guidance in the selection of methods by – prescribing a reduced set of effective methods that optimize the selection criteria, and facilitating selection among these methods through documentation of their pros and cons.

## 4. CONCLUSION

Our research has two primary objectives – (1) to overcome the limitations of the current requirements engineering models and (2) to enhance the limited guidance in selecting methods for activities in the requirements generation process. We propose a model that addresses the complete requirements process and which identifies activities at an appropriate level of activity abstraction. In contrast, current models either have a narrow focus or provide a high-level perspective of the requirements process. The decomposition in our model illustrates a better evolutionary path of the requirements. In addition, it also facilitates the mapping of methods to activities. We also identify the objectives of the activities and explicitly state them. The identification of objectives is imperative because the mapping of methods to activities is driven by the objectives. In addition to overcoming the limitations of current models, our proposed model is novel in that it generates quality requirements in a cost-effective manner. This is achieved by performing the V&V activities iteratively on smaller sets of requirements as they relate to individual system functionalities.

To enhance the guidance in method selection, we have identified methods that are commonly used in the industry and have mapped them to activities based on activity objectives. Thus, we achieve a synchronization of methods and activity objectives, unlike current research that maps methods to high-level activity objectives. To further simplify the task of method selection, we have identified four selection criteria (cost, time, personnel, completeness) that are widely used in the industry and have grouped those methods based on these criteria. As a result, for each activity we have identified a much smaller set of methods that optimize each of the four selection criteria. This setting enhances the guidance in selecting the most appropriate method for an activity based on a selection criterion or a combination of criteria. In effect, we have provided a framework that enables the requirements engineer to make informed decisions in the method selection process.

The benefits attributed to our model are substantiated through literature citation and rationalization based on experiences in the industry. In the future, we envision a detailed empirical evaluation to provide better insights into the implementation aspects of the expanded RGM model and our approach. In addition, continuing the mapping of methods for the remainder of the software development life cycle can provide additional guidance for the software engineering community, and subsequently improve the success rate of software projects.